\documentclass[journal]{IEEEtran}
\usepackage{amsmath, amssymb, bm, cite, epsfig, psfrag}
\usepackage{graphicx}
\usepackage[margin=0.7in]{geometry}
\usepackage{dblfloatfix}
\usepackage{array}
\usepackage{lipsum}
\usepackage{ragged2e}
\newcolumntype{P}[1]{>{\centering\hspace{0pt}}p{#1}}
\newcolumntype{M}[1]{>{\centering\hspace{0pt}}m{#1}}
\newcolumntype{L}{>{\centering\arraybackslash}m{3cm}}
\usepackage[font=small]{caption}
\usepackage{epstopdf}	
\usepackage{longtable}
\usepackage{supertabular,booktabs}
\usepackage{bbm}
\usepackage{multirow}
\usepackage[usenames,dvipsnames]{xcolor}

\usepackage{subfigure}
\usepackage{etoolbox}
\usepackage{pbox}
\usepackage{fixltx2e}
\usepackage{filecontents}
\usepackage{enumerate}

\usepackage{colortbl}
\usepackage{fancyhdr}

\usepackage{bm}
\newtoggle{conference}
\togglefalse{conference} 
\graphicspath{{figures/}}

\usepackage[marginal]{footmisc}

\fancypagestyle{firststyle}{
	\fancyhf{}
	\fancyhead[C]{Conference on}     
	\fancyfoot[L]{This is a notice}

}

\fancypagestyle{firststyle}{
	\fancyhf{}
	\fancyhead[C]{S. Ju and T. S. Rappaport, ``Simulating Motion - Incorporating Spatial Consistency into the NYUSIM Channel Model,'' \textit{2018 IEEE 88th Vehicular Technology Conference Workshops (VTC2018-Fall WKSHPS)}, Chicago, USA, Aug. 2018, pp. 1-6.}     

}

\begin{document}
\bibliographystyle{IEEEtran}
\title{Simulating Motion - Incorporating Spatial Consistency into the NYUSIM Channel Model}
\author{\IEEEauthorblockN{Shihao Ju and Theodore S. Rappaport\\}
\IEEEauthorblockA{	\small NYU WIRELESS\\
					NYU Tandon School of Engineering\\
					Brooklyn, NY 11201\\
					\{shao, tsr\}@nyu.edu}}
\maketitle
\thispagestyle{firststyle}
\footnote{This work is supported in part by the NYU WIRELESS Industrial Affiliates, and in part by the National Science Foundation under Grants: 1702967 and 1731290.}
\begin{abstract}
This paper describes an implementation of spatial consistency in the NYUSIM channel simulation platform. NYUSIM is a millimeter wave (mmWave) channel simulator that realizes measurement-based channel models based on a wide range of multipath channel parameters, including realistic multipath time delays and multipath components that arrive at different 3-D angles in space, and generates life-like samples of channel impulse responses (CIRs) that statistically match those measured in the real world \cite{Sun17b}. To properly simulate channel impairments and variations for adaptive antenna algorithms or channel state feedback, channel models should implement spatial consistency which ensures correlated channel responses over short time and distance epochs. The ability to incorporate spatial consistency into channel simulators will be essential to explore the ability to train and deploy massive multiple-input and multiple-output (MIMO) and multi-user beamforming in next-generation mobile communication systems. This paper reviews existing modeling approaches to spatial consistency, and describes an implementation of spatial consistency in NYUSIM for when a user is moving in a square area having a side length of 15 m. The spatial consistency extension will enable NYUSIM to generate realistic evolutions of temporal and spatial characteristics of the wideband CIRs for mobile users in motion, or for multiple users who are relatively close to one another.

\end{abstract}
    

\section{Introduction}~\label{sec:intro}
There is growing interest in emerging ultra-wide bandwidth and low latency communications \cite{Fatema17a}. Mmwave mobile communication has the potential to satisfy such requirements, and ongoing measurements and modeling are aimed to understand the characteristics of mmWave channels that can support unprecedented bandwidths \cite{Rap13a}. 

The third generation partnership project (3GPP) conceived of a system-level drop-based statistical modeling approach to generate spatial CIRs. The 3GPP channel models \cite{3GPP.38.901}\cite{Sun18a} describe clusters by a joint delay-angle probability density function so that a group of traveling multipaths that have close excess delays must depart and arrive from similar angles of departure (AODs) and angles of arrival (AOAs). The 3GPP Release 14 channel model uses a fixed number of clusters and subpaths in each cluster. For instance, in the non-line-of-sight (NLOS) urban microcell (UMi) environment, the number of clusters is 19, and the number of subpaths in each cluster is 20 \cite{3GPP.38.901}\cite{Sun18a}, which is not supported by the real-world measurements at mmWave bands \cite{Sun17b}\cite{Sun18a}\cite{Samimi15a}.

Spatial consistency, which means a user terminal (UT) will experience a similar scattering environment which causes smooth channel transitions due to the UT motion, has been acknowledged in the Release 14 3GPP standards \cite{3GPP.38.901}. For new 5G applications such as vehicle-to-vehicle (V2V) communications \cite{Perfecto17a} and Internet of Things (IoT) \cite{Kong17a}, a spatially consistent channel model which can accurately characterize the temporal and spatial evolution of channel impulse responses (CIRs) over a local area is necessary. 3GPP proposes two alternative UT mobility procedures for spatial consistency and provides some insights for further spatial consistency research \cite{3GPP.38.901}.
\begin{figure*}[]
	\setlength{\abovecaptionskip}{0.cm}
	\setlength{\belowcaptionskip}{-0.5cm}
	\centering
	\includegraphics[width=0.9\textwidth]{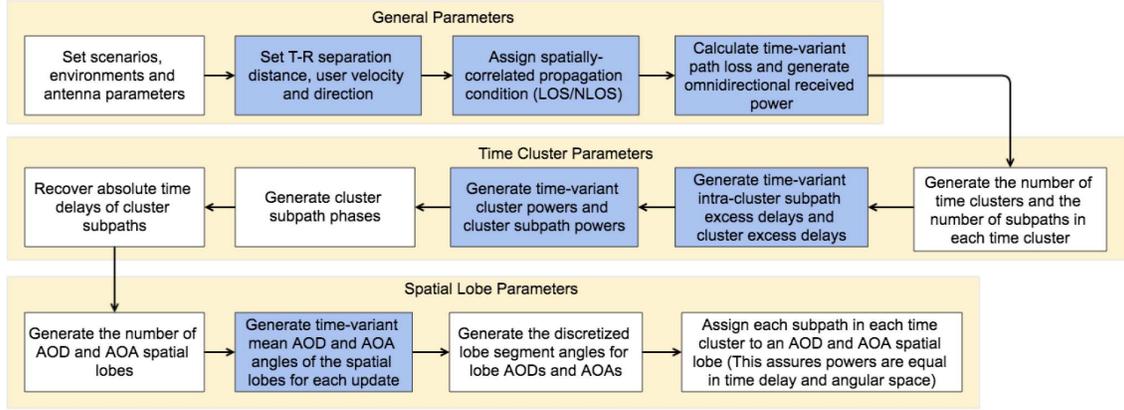}
	\caption{Spatially consistent channel coefficient generation procedure}
	\label{fig:sc_v1}
\end{figure*}

NYUSIM is an open source simulator based on a 3-D spatial statistical channel model (SSCM) \cite{Sun17b}\cite{Sun18a}\cite{Samimi15a}, which adopts a time-cluster spatial-lobe (TCSL) approach to characterize temporal and angular properties of multipath components. NYUSIM provides physical-level simulations that realistically simulate the downlink channel between a base station (BS) and a UT. The output files from NYUSIM include omnidirectional and directional CIRs, and joint AOA and AOD power spectrum. NYUSIM works over a wide range of frequencies between 800 MHz and 100 GHz and provides CIRs for many types of antenna arrays for MIMO systems \cite{Samimi16d}. To date, NYUSIM, as an open source software, has been downloaded more than 50,000 times from industry and the academic research community. 

A new spatial consistency feature can extend the NYUSIM channel model from a static drop-based model (a static CIR at a particular distance/location, with random draws for each successive mobile location) to a dynamic model that first creates an anchor CIR and then builds realistic time and distance-varying CIR samples that have correlated statistics, as a UT moves over a local area. An early bit error rate simulator (BERSIM) \cite{Fung93a} developed for the measurement-based statistical indoor CIR model (SIRCIM) \cite{Rap91b} showed that BER was not only a function of root mean square (RMS) delay spread but was also a function of the temporal and spatial distribution of multipath components as the UT moves. Thus, evaluating system performance should take spatial consistency into account. 

This paper extends NYUSIM with spatial consistency to consider the impact of track motion on CIRs over distances of up to 15 m, as suggested in 3GPP Release 14 \cite{3GPP.38.901}, where 15 m represents a time epoch of 10 s when a UT travels at a walking speed of 5 km/h. Specifically, NYUSIM uses velocity and the direction of UT motion as input information and generates CIRs as a function of time and distance by making large-scale and small-scale parameters spatially correlated random variables.  

This paper proposes a framework of the modified channel coefficient generation procedure for spatial consistency in NYUSIM, and is organized as follows. Section \ref{sec:previous} presents available schemes for spatial consistency based on stochastic or deterministic models.  In Section \ref{sec:nyusim}, we investigate the effects of track motion on the parameters such as the number of time clusters and spatial lobes, and introduce modified procedures to ensure spatial consistency in the NYUSIM channel model. Conclusions and insights for future work are presented in Section \ref{sec:conclusion}.

\section{Early Research on Spatial Consistency}\label{sec:previous}
For 5G requirements and applications, conventional drop-based modeling is inadequate \cite{Shafi18a}. It is essential to understand spatial consistency and correlation of channels between two close-by locations for two different UTs, or for a UT that is traveling in a local area. If a UT is moving in a local area or if multiple UTs are very closely spaced, their scattering environment should be similar. 
Thus, closely spaced channel snapshots should have similar delays, AOAs and AODs \cite{METIS15a}. Such modeling was first implemented in SIRCIM and SMRCIM for indoor and outdoor channels in the 1990s, before the use of directional antennas \cite{Seidel90a}\cite{Nuckols99}.

For directional mmWave wireless communications, the independent generation of channel parameters for closely spaced UT positions from a fixed probabilistic distribution is not appropriate. Therefore, spatially consistent models are required to capture smooth transitions of channel characteristics in a dynamic way over relatively closely spaced locations. 
In other words, the large-scale and small-scale parameters used in the CIR generation are assumed to be consistent and correlated within a grid having a side length of approximately 15 m \cite{3GPP.38.901}. In the NYUSIM channel model, the large-scale parameters are the shadow fading, the number of time clusters, the number of spatial lobes; the small-scale parameters are the random variables with respect to excess time delays, power fading, and angular spreads. Each parameter obeys different statistical distributions\cite{Sun17b}\cite{Sun18a}\cite{Samimi16a}. Another concept that may be confused with spatial consistency is the small-scale spatial statistics that are small-scale fading and small-scale autocorrelation of the received voltage amplitudes \cite{Rap17b}. The rapid decorrelation of received voltage amplitudes occurred over 0.67 to 33.3 wavelengths (0.27 cm to 13.6 cm) at 73 GHz \cite{Rap17b}. It is important to note that spatial consistency does not deal with the small-scale correlation of received power levels, but rather focuses on providing a consistent and correlated scattering environment that the moving UT experiences. 

\subsection{3GPP TR 38.901 Models}
3GPP introduced two alternative procedures for spatial consistency\cite{3GPP.38.901}. The first approach is to update channel cluster power/delay/angles at time $t_k$ based on the power/delay/angles, UT speed vector, and UT position at time $t_{k-1}$ following the equations in \cite{3GPP.38.901}. 
In the second approach, channel parameters such as delay/power/angles are still independent for different UT positions, but the generation procedure of these parameters is modified. Specifically, the modified steps generate delays and angles based on uniform distributions where the coefficients linearly depend on the correlation distance to assure spatial consistency of the simulated channel over a track \cite{3GPP.38.901}. Correlation distance is the distance beyond which the auto-correlation value of a large-scale parameter falls below 0.5 \cite{Wang16a}, and varies according to different large-scale parameters such as angular spread, delay spread, K-factor and shadow fading, and typically ranges from 12 m to 15 m in UMi environment \cite{3GPP.38.901}. Note that in \cite{Rap17b}, there was a high correlation of averaged received power levels over grids with 5 m length, even though directional antennas showed the small-scale correlation of received power was less than a meter and depended on the directional pointing angle. An experiment following this first method where a UT moved at a speed of 0.8333 m/s with the update distance 0.1 m showed that the cluster delays and angles fluctuate continuously according to the UT trajectory \cite{Shafi18a}, and \cite{Rap17b} showed similarly that the antenna beam pointing direction impacts the small-scale correlation behavior of the received voltage envelope. 

\subsection{WINNER II Models}
As proposed in the WINNER II channel models \cite{WinnerII}, a channel segment represents a quasi-stationary period during which the change of probability distributions of large-scale parameters can be ignored. Based on channel segments and drops, WINNER II implements spatial consistency by characterizing the transition between segments. The clusters of ``old'' segments are replaced by the clusters of ``new'' segments, and the powers received in the old segments ramp down, and the powers received in the new segments increase. A more advanced method is based on a Markov process, which requires reliable parameters that are not available from current measurements \cite{WinnerII}. This Markov method, nevertheless, offers promise since it is capable of providing the most realistic behavior of the channel among these methods.

\subsection{5GCM Models}
\begin{figure}[]
	\setlength{\abovecaptionskip}{0.cm}
	\setlength{\belowcaptionskip}{-0.5cm}
	\centering
	\includegraphics[width=0.45\textwidth]{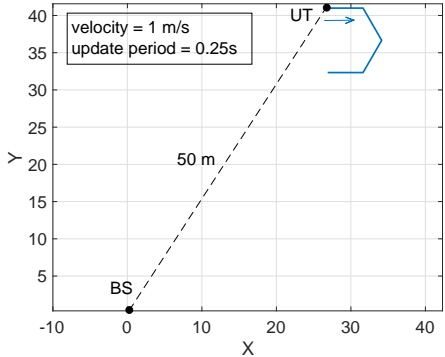}
	\caption{UT trajectory. The T-R separation distance is 50 m, the total walking distance is 20 m with the velocity 1 m/s, half-hexagon route. The update distance is 0.25 m. }
	\label{fig:user_trajectory}
\end{figure}
Three geometry-based stochastic modeling options for spatial consistency are presented in \cite{A5GCM15}. The first option adds spatial consistency to cluster-specific random variables, LOS/NLOS, and indoor/outdoor states \cite{3GPPTDOC161622}. The algorithm generates a grid for cluster-specific random variables, the granularity of which is the parameter-specific correlation distance. These cluster-specific random variables include time excess delay, shadowing, angle offset and sign, which are generated by interpolating four complex normal distributed independent and identical distribution (i.i.d) random variables on four vertices of one grid. Results \cite{Ademaj17a} suggest that this 5GCM applied interpolation method is not appropriate because the correlated values are different for equal distances from an anchor point, and they introduced a bilinear interpolation approach to generate cluster-specific random variables. 

The second approach is described in \cite{Wang16a} called geometry stochastic approach. The general idea is that the large-scale and small-scale parameters evolve continuously and predictably over a local area. Large-scale parameters are shared among close UTs with small variations instead of the random and independent generation of the parameters for each UT at different positions in a local area. To characterize dynamic variations of small-scale parameters, this geometry stochastic approach introduces time-variant angles and cluster birth and death over a given time epoch. This second method uses a low-complexity linear approximation to estimate the variant spatial angles including AOA, AOD, zenith angle of arrival (ZOA), and zenith angle of departure (ZOD) based on the angle information of the UT at time $t$. Also, cluster birth and death is characterized as a Poisson process where the cluster birth and death always occur simultaneously because of the fixed number of clusters assumed in \cite{A5GCM15}. 


The third approach uses geometric locations of clusters and is introduced in \cite{A5GCM15}, where cluster and path angles and delays are translated into geometrical positions of the corresponding scatterers. Each UT has a circular area with a radius of the correlation distance of large-scale parameters. If two users' circles do not overlap, they are assumed to be independent. If a UT is located within another UT's circle, only the strongest few clusters of both users will be used to generate channel coefficients. The clusters may be calculated simply at the closest grid points around the UT position to avoid excessive use of memory \cite{A5GCM15}. 

\begin{figure}[]
	\setlength{\abovecaptionskip}{0.cm}
	\setlength{\belowcaptionskip}{-0.5cm}
	\centering
	\includegraphics[width=0.4\textwidth]{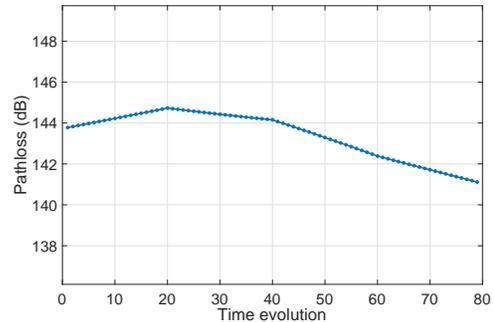}
	\caption{Time-variant path loss, calculated by the pathloss equations from \cite{Sun18a}, in the UMi scenario with the half-hexagon route in Fig. \ref{fig:user_trajectory}. }
	\label{fig:pathloss}
\end{figure}

\subsection{COST 2100 Models}
The COST 2100 channel model is a geometry-based stochastic channel model (GSCM) that can reproduce the stochastic properties of radio channels over time, frequency, and space \cite{COST2100}. The visibility region as defined in \cite{COST2100} is a key concept of geometric and stochastic propagation models, and describes the space- or time-span over which a group of traveling multipath components are visible at a UT \cite{Verdone12a}. The environment is first generated using COST 2100 models. The motion of the UT in the simulated area causes the visibility of different clusters to change as the UT enters and leaves different visibility regions. Thus, spatial consistency can be modeled as the evolution of visibility regions.

\subsection{Map-based Models}
Map-based channel models are compared with geometry-based stochastic channel models in realizing spatial consistency \cite{Turkka16a}. Spatial consistency is assessed by analyzing the signal-to-interference ratio (SIR) of a multiuser multiple-input and single-output (MU-MISO) system. In a spatially consistent channel model, where close-by UTs are expected to have highly correlated channels, the MU-MISO SIRs are low. Results suggest that a map-based approach estimates the MU-MIMO system performance more accurately in both LOS and NLOS environments \cite{Turkka16a}. 
It cannot be ignored that map-based channel models may introduce complex channel generation mechanism and cause high computation complexity \cite{Molisch16a}.

\subsubsection{MiWEBA Models}
Deterministic or ray-tracing techniques are also used to develop channel models such as mmWave evolution for backhaul and access (MiWEBA) \cite{MiWeba14a}. MIWEBA is a quasi-deterministic channel model for mmWave communication links at the 60 GHz band. MiWEBA adopts ray tracing techniques and sums up a few strong deterministic paths and several relatively weak random rays. The deterministic rays are obtained from path-length geometry between the transmitter (TX) and receiver (RX), and signal power is computed using distance and Friis' free space path loss equation, while weak rays are sampled from measurement-based statistical distributions. The spatial consistency is inherently in the MiWEBA model since a few strong multipath components are computed deterministically. Thus, generated CIRs at two close locations are similar and highly-correlated from the ray tracing of the physical environment.
\begin{figure}[]
	\setlength{\abovecaptionskip}{0.cm}
	\setlength{\belowcaptionskip}{-0.5cm}
	\centering
	\includegraphics[width=0.5\textwidth]{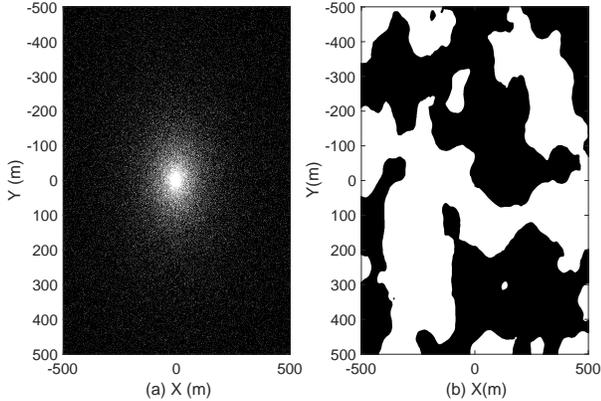}
	\caption{The UMa (a) uncorrelated and (b) correlated LOS/NLOS conditions in a 500 m $\times$ 500 m square area based on the LOS probability from \cite{3GPP.38.901}.}
	\label{fig:los}
\end{figure}
\subsubsection{METIS Models}
Mobile and wireless communications Enablers for the Twenty-twenty Information Society (METIS) \cite{METIS15a} is a hybrid model that combines map-based and stochastic models. METIS uses ray-tracing techniques to obtain large-scale parameters for a site-specific environment, and measurement-based results to model the small-scale parameters. The spatial consistency is also inherently incorporated within the METIS model because the generated large-scale parameters of two close locations are almost the same, since ray tracing of the physical environment is used to simulate the spatial and time evolution.

\section{Spatial Consistency for NYUSIM} \label{sec:nyusim}

Spatial consistency requires the channel realizations for the large-scale and small-scale parameters to vary continuously and realistically as a function of UT location. Under the framework of the NYUSIM channel model \cite{Sun17b}\cite{Sun18a}\cite{Samimi16a}, we propose a dynamic extension for spatial consistency shown in Fig. \ref{fig:sc_v1}. The steps in the blue blocks of the flowchart are modifications to realize spatial consistency. The main challenge here is to extend a static and drop-based channel model to a dynamic and time-variant one. The simulator will update channel coefficients in an iterative way along the UT trajectory. Each modified step is described as follows.

\begin{enumerate}
\item \textit{Set T-R separation distance, UT velocity, and direction:} Instead of generating T-R separation distance randomly in the distance range, the distance, UT velocity, and direction can be set before the simulations. The update distance of UT is less than 1 m so that we can capture small variations compared to the 15 m correlation distance of large-scale parameters. For illustration purpose, the update period is 0.25 s when the UT velocity is 1 m/s in the following simulations, which means that there are 80 points over a distance of 20 m and a time epoch of 20 s. The simulation sample of UT trajectory is shown in Fig. \ref{fig:user_trajectory}. The black dotted line connects BS and UT. The blue arrow shows the UT direction. 

\item \textit{Assign spatial-correlated propagation condition (LOS/NLOS):} Since the available NYUSIM works for drop-based simulations, the LOS/NLOS does not change during the procedure. A moving UT may experience different propagation conditions in real life, and in a local area, the propagation conditions at multiple positions are correlated. Therefore, we follow a procedure similar to the one described in WINNER II \cite{WinnerII} where the model generates the correlation of large-scale parameters by applying an exponential spatial filter to the independent random values.
\begin{equation}
\tilde{r}_{m,n} = \sum^M_{x=0}\sum^N_{y=0}r_{m,n}h(m-x,n-y)
\end{equation}
, where $r_{m,n}$ and $\tilde{r}_{m,n}$ are independent and correlated random variables at position $(x,y)$, respectively. $M\times N$ is the total number of the grid points and $h$ is the exponential filter which is defined as 
\begin{figure}[]
	\setlength{\abovecaptionskip}{0.cm}
	\setlength{\belowcaptionskip}{-0.5cm}
	\centering
	\includegraphics[width=0.5\textwidth]{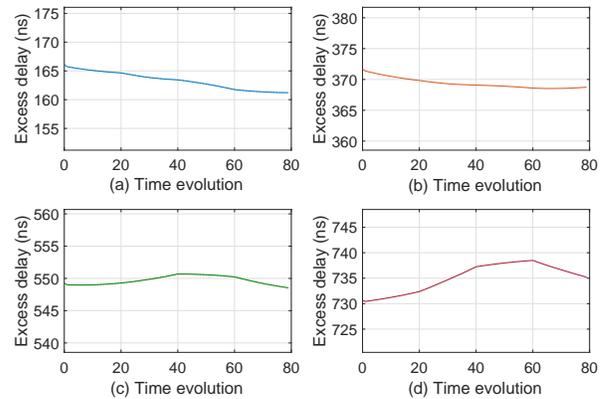}
	\caption{Cluster excess delays of the four NLOS clusters in the UMi scenario with the half-hexagon route in Fig. \ref{fig:user_trajectory}.}
	\label{fig:excessdelay}
\end{figure}
\begin{equation}
h(m,n) = exp\left(-\frac{\parallel m-n \parallel}{\Delta d}\right)
\end{equation} 
where $\parallel m-n \parallel$ is the distance between UT and BS, $\Delta d$ is the side length of grid and the corresponding correlation distance that is 50 m for UMi and UMa scenarios \cite{3GPP.38.901}. The uncorrelated and correlated LOS/NLOS conditions are shown in Fig. \ref{fig:los}. 

\item \textit{Calculate time-variant path loss}: The path loss is calculated from the updated UT position in each update. The shadow fading standard deviation can be regarded as a constant in a local area and is changed only when the LOS/NLOS condition changes. For the route shown in Fig. \ref{fig:user_trajectory}, the corresponding path loss variations are shown in Fig. \ref{fig:pathloss}. As expected, the path loss first increases and then decreases since the UT moves in a half-hexagon route.

\item \textit{Generate time-variant cluster and subpath excess delays:} The initial subpath and cluster delays are generated by the original NYUSIM method where the UT is at the starting point of a half-hexagon route. As the UT moves along the path, time excess delays are updated based on the similar procedure in \cite{3GPP.38.901} where the update is based on the velocity, direction, and the time-variant AOAs and AODs. Note that the excess delay depends on the angular information at the most recent moment. The cluster excess delays, which are the delays of first subpaths in all clusters, are shown in Fig. \ref{fig:excessdelay}. Notice that the delays experience a smooth evolution instead of a random variation. The spatial consistency of time clusters in a local area is thereby provided based on the most recent update. Note that for the LOS cluster, the first subpath is LOS component whose excess delay is zero.

\item \textit{Generate time-variant cluster powers and subpath powers:}
Because of the time-variant (or distance-variant) path loss, the total received power also varies. Further, the power of each subpath in each cluster needs to be allocated during the update in a correlated and continuous way. The two key parameters in this step are the shadowing factor (SF) for a cluster, and the SF for a subpath, which are characterized as lognormal random variables with zero dB means, and $\sigma_Z$ and $\sigma_U$ standard deviations, respectively. Here, a similar exponential spatial filter to the one used to ensure correlated LOS probability is applied to make the SFs correlated. The uncorrelated and correlated values of SFs at 73 GHz with 1 m/s speed and 0.25 s update period are shown in Fig. \ref{fig:power}. The figure shows that the variations of power-related parameters are smoother after the filter. The chosen correlation distance is 15 m, which refers to the value from 3GPP \cite{3GPP.38.901}, but more accurate values of these parameters will be needed and measured for different environments.  
\begin{figure}[]
	\setlength{\abovecaptionskip}{0.cm}
	\setlength{\belowcaptionskip}{-0.5cm}
	\centering
	\includegraphics[width=0.5\textwidth]{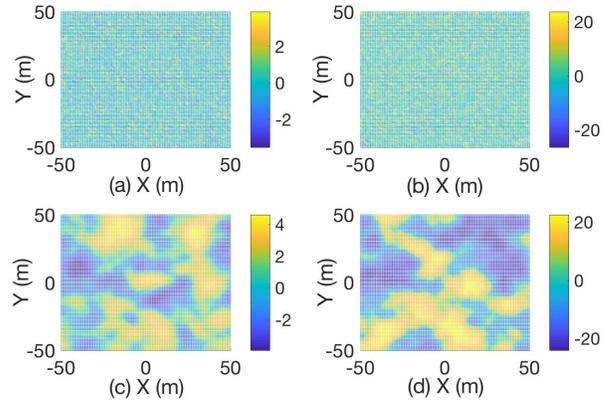}
	\caption{(a) Uncorrelated $\sigma_Z$, (b) Uncorrelated $\sigma_U$, (c) Correlated $\sigma_Z$, (d) Correlated $\sigma_U$ power-related parameters predefined in a 50 m $\times$ 50 m grid map. }
	\label{fig:power}
\end{figure}
\item \textit{Generate time-variant subpath AODs and AOAs:} One critical principle for the implementation of spatial consistency is to minimize computation complexity. Most of the spatial consistency algorithms generate grids for many parameters, which suffer from high computation complexity. Furthermore, it is not easy to compute accurate AODs and AOAs for NLOS multipath components. 

A linear approximation for angles was introduced in \cite{Wang16a}, which is written as
\begin{equation}
\theta_{n,Angle}(t)=\theta_{n,Angle}(t_0)+S_{n,Angle}(t-t_0)
\end{equation}
where $n$, $m$, $Angle$ are the $n$-th time cluster, the $m$-th subpath in the $n$-th time cluster, and $Angle$ is an index for AOD, ZOD, AOA, and ZOA. Here, the changing rate of angles can be seen as a slope $S_{n,m,Angle}$. For LOS clusters, the expressions of slopes are given by \cite{Wang16a}
\begin{eqnarray}
S_{n,AOD}=&\frac{v\sin(\phi_v(t_0)-\phi_{n,AOD}(t_0)+\psi_{n,AOD})}{d_{2D}}\\
\label{eqn:ang4}
S_{n,ZOD}=&-\frac{v\cos(\phi_v(t_0)-\phi_{n,AOD}(t_0)+\psi_{n,ZOD})}{d_{3D}}\\
S_{n,AOA}=&-\frac{v\sin(\phi_v(t_0)-\phi_{n,AOA}(t_0)+\psi_{n,AOA})}{d_{2D}}\\
S_{n,ZOA}=&-\frac{v\cos(\phi_v(t_0)-\phi_{n,AOA}(t_0)+\psi_{n,ZOA})}{d_{3D}}
\label{eqn:ang7}
\end{eqnarray}

where $d_{2D}$ is the azimuth distance between BS and UT (without considering antenna heights); $d_{3D}$ is the true 3-D distance. $\psi_{n,AOD}$, $\psi_{n,AOA}$, $\psi_{n,ZOD}$, $\psi_{n,ZOA}$ are cluster specific reflection surface angles which are modeled as uniform random variables. The AODs and AOAs are generated uniformly from $(-\pi,\pi]$ while the ZODs and ZOAs $(-\pi/2,\pi/2]$ for NLOS clusters. For LOS clusters, all four reflection angles in Eqn. \ref{eqn:ang4}-\ref{eqn:ang7} are simply zero.

The simulation results of angular information for 73 GHz UMi scenario with same settings as above are shown in Fig. \ref{fig:angless}, which shows that the only line that is not ``shaking'' is the LOS component. Other NLOS components have some random variations over 20 s simulation of motion as the UT moves at 1 m/s. The angles are changing smoothly as the UT moves. 

\end{enumerate}

The preceding six steps provide a viable means for generating the appropriate statistics for excess delays, powers, and angles to include spatial consistency to NYUSIM. The simulation results show that the CIRs generated from NYUSIM experience a smooth evolution along the UT trajectory, which also matches the physical expectation. Cluster birth and death is an interesting concept that is not considered in the current extended model. Cluster birth and death introduces obvious power fluctuations, and the corresponding delays and angles will change, as well. The rate of cluster birth and death is site-specific. While early work offers strong evidence \cite{Rap17b}, more work can be done to capture the dynamic characteristics of mmWave radio channels. 

\begin{figure}[]
	\setlength{\abovecaptionskip}{0.cm}
	\setlength{\belowcaptionskip}{-0.5cm}
	\centering
	\includegraphics[width=0.5\textwidth]{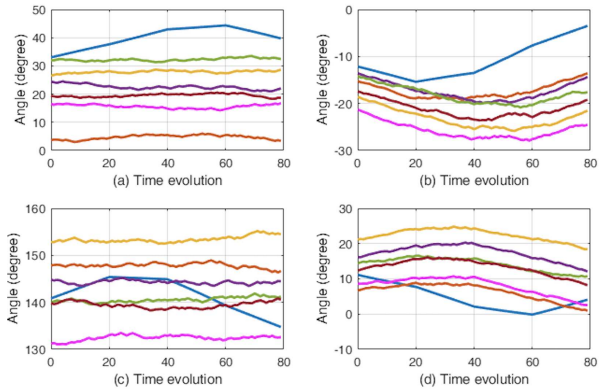}
	\caption{ (a) AOD, (b) ZOD, (c) AOA, (d) ZOA time-variant angles after linear approximation with the half-hexagon route in Fig. \ref{fig:user_trajectory}. }
	\label{fig:angless}
\end{figure}

\section{Conclusion}\label{sec:conclusion}
This paper develops a new channel model extension of spatial consistency for NYUSIM and builds upon approaches and algorithms for other channel models. 
The model extension given here enables the simulator, NYUSIM, to implement CIRs with UT motion while maintaining realistic correlations on small- and large-scale parameters. This extension provides guidance for the strategies of channel modeling when considering multiuser beamforming and beam alignment. Since the NYUSIM channel model is statistical, it is reasonable to maintain some randomness in the simulation results. Without losing the original framework of NYUSIM and the concepts of time clusters and spatial lobes, the extended channel model can deal with dynamic conditions using spatial consistency.

\bibliographystyle{IEEEtran}
\bibliography{vtc}

\end{document}